\documentclass[prl,twocolumn,superscriptaddress]{revtex4-1}

\usepackage{graphics,graphicx}

\def\refjpa#1#2#3{J.~Phys.~A,~{\bf #1},\ #2,\ (#3)}

\def\pr#1#2#3{Phys.~Rev,~{\bf #1},\ #2,\ (#3)}
\def\pra#1#2#3{Phys.~Rev.~A,~{\bf #1},\ #2,\ (#3)}

\def\prl#1#2#3{Phys.~Rev.~Lett.,~{\bf #1},\ #2,\ (#3)}

\def\etal{{\it et al.}\,}

\def\bpmat{\begin{pmatrix}}
\def\epmat{\end{pmatrix}}
\def\bmat{\begin{matrix}}
\def\emat{\end{matrix}}

\def\1{\mbox{1\hskip-.25em l}}

\def\beq{\begin{equation}}
\def\eeq{\end{equation}}
\def\bw{\begin{widetext}}
\def\ew{\end{widetext}}
\def\beqar{\begin{eqnarray}}
\def\eeqar{\end{eqnarray}}

\def\scheq{Schr\"{o}dinger\, equation\,}

\begin{document}
\title{\bf The absolute position of a resonance peak}


\author{Shachar Klaiman}
\email{shachark@technion.ac.il}
\affiliation{Schulich Faculty
of Chemistry and Minerva Center for Nonlinear Physics of Complex
Systems, Technion -- Israel Institute of Technology, Haifa 32000,
Israel.}
\author{Nimrod Moiseyev}
\affiliation{Schulich Faculty
of Chemistry and Minerva Center for Nonlinear Physics of Complex
Systems, Technion -- Israel Institute of Technology, Haifa 32000,
Israel.}
\affiliation{The Department of Physics, Technion -- Israel Institute of Technology, Haifa 32000,
Israel.}

\begin{abstract}
It is common practice in scattering theory to correlate between the position of a resonance peak in the cross section and the real part of a complex energy of a pole of the scattering amplitude. In this work we show that the resonance peak position appears at the absolute value of the pole's complex energy rather than its real part. We further demonstrate that a local theory of resonances can still be used even in cases previously thought impossible.
\end{abstract}
\maketitle

There are few phenomena in physics which are as striking and ubiquitous as resonance peaks in the cross section. As such, it is of great importance to understand their origin and to be able to predict the resulting resonance profile in the cross section. Over the years many methods have been proposed and successfully used to analyze and explain resonance peaks appearing in the cross section. Among which we mention the use of the poles of the scattering matrix(S-matrix) \cite{newton}, the Fano configuration interaction \cite{fano}, and the Feshbach resonance theory \cite{feshbach}.

The main purpose of such theories is to reconstruct the resonance structure appearing in the cross section. This should be done with but a few parameters which can be used to characterize the scattering system at hand. It is in the way these parameters are chosen and calculated that the methods mentioned differ from each other. One such difference is the use of non-local versus local parameters, i.e., whether the system parameters used to describe the resonance profile are energy dependent or not.

The most famous application of a local resonance theory is the Breit-Wigner(BW) resonance profile \cite{Breit-Wigner} which connects an appearing resonance peak in the cross section with a pole of the S-matrix in the complex energy plane. The resulting resonance profile reads:
\beq
\label{eq:Breit-Wigner}
\sigma(E)\propto\frac{\left(\Gamma/2\right)^2}{\left(E-\epsilon\right)^2+\left(\Gamma/2\right)^2},
\eeq
where, $\epsilon$ is known as the resonance position and is given by the real part of the pole in the energy plane, and $\Gamma$ is width of the Lorentzian at half maximum which is twice the imaginary part of the resonance pole in the energy plane. Although the above resonance profile has a very impressive track record it fails whenever the pole is not isolated (from the effect of other poles), it is near the threshold, or it is far from the real axis. One possible resolution to this predicament is to move into a non-local description and define and energy dependent width and position such that the resonance profile now reads, $\sigma(E)\propto\frac{\left(\Gamma(E)/2\right)^2}{\left(E-\epsilon(E)\right)^2+\left(\Gamma(E)/2\right)^2}$.
Although clearly able to reproduce any resonance structure it is neither straight forward nor simple to calculate the non-local terms in the above equation for an arbitrary scattering potential.

\begin{figure}[htbp]
\includegraphics[width=\columnwidth,angle=0,scale=1,draft=false,keepaspectratio=true]{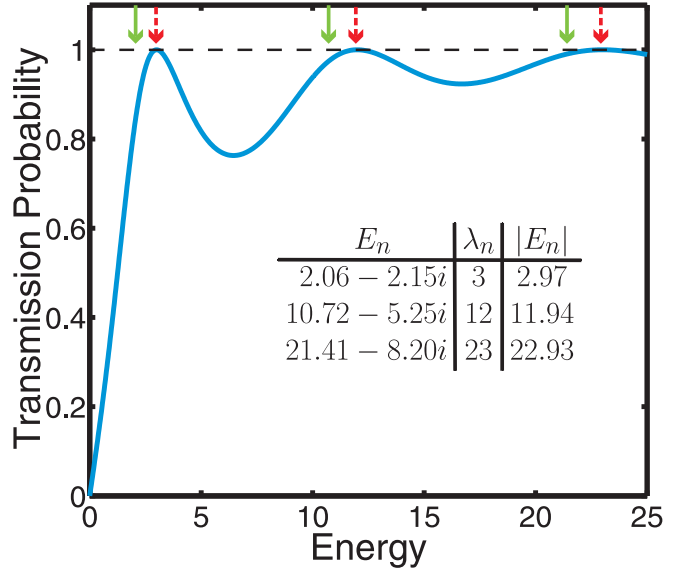}
\caption{(color online) The transmission probability above a square well potential of depth $V_0=-13$ and length $L=\pi/\sqrt{2}$. Full(green) and dashed(red) arrows mark the real part and absolute value of the resonance poles energy respectively. The table in the inset shows the values of the first three energies, $E_n$, of the resonance poles, the analytical position, $\lambda_n$, of the transmission unities, and the absolute value of the resonance energies.}
\label{FIG::sqwell}
\end{figure}

Recently, the use of the poles of the scattering matrix was reinvigorated by the novel method presented by Tolstikhin \etal \cite{tolstikhinPRL}. The possibility to solve the \scheq  with Siegert boundary conditions \cite{Siegert} to obtain all the poles of the scattering matrix opened the door to describe the entire cross section using only the poles. Tracing the origin of the structures in the cross section to specific poles of the matrix has always been invaluable in order to understand the scattering mechanisms responsible for the resulting cross section. It would, therefore, be extremely beneficial if one could assign to every resonance in the cross section a specific pole. It is the failure of this possibility that necessitates the use of non-local theories and as we shall demonstrate there is still much to be done using only local resonance profiles.

In order to demystify the above discussion let us start with the simplest example where one cannot connect between the poles of the S-matrix and the resonance appearing in the cross section using the assumptions of the BW profile. Consider a one dimensional square well. It is well known, see for example \cite{Griffiths}, that the transmission spectrum displays an oscillating behavior and the transmission probability reaches unit value only at discrete energies satisfying: $\lambda_n=-V_0+\frac{\hbar^2\pi^2n^2}{2mL^2}$,
such that $T(\lambda_n)=1$. In the above equation $V_0$ and $L$ are the well's depth and length respectively, $m$ is the mass of the projectile, and $n$ is a positive integer. Fig. \ref{FIG::sqwell} depicts the transmission for a specific choice of the wells depth and length. The poles of the scattering amplitude can be calculated by solving the \scheq with outgoing boundary conditions \cite{zavin}. The positions of the first three resonance poles, i.e., real part of the complex energy of the poles appearing in the fourth quadrant, are depicted in Fig. \ref{FIG::sqwell} using full(green) arrows. As can be readily observed one cannot directly assign a transmission unity with the real part of the resonance pole's energy. At first this may seem obvious since the resonance poles clearly overlap no direct connection between a single pole and the observed resonance can be expected. But, as we shall show in this letter there is a one to one correspondence between each transmission unity and a single resonance pole. To relieve the suspense we give here the final answer that the resonance positions in the cross section are related to the absolute value of the pole's complex energy rather than to its real part. This is demonstrated in Fig. \ref{FIG::sqwell} by the dashed(red) arrows indicating the absolute values of the complex resonance poles' energies.

Let us consider a scattering process from a single one dimensional channel. As mentioned above, the transmission amplitude can be expressed using \textit{all} the poles of the scattering matrix \cite{tolstikhinPRL}. These are found as solutions of the \scheq: $\left[-\frac{1}{2}\partial^2_x+V(x)\right]\Psi_n(x)=\frac{k_n^2}{2}\Psi_n(x)$ along with Siegert boundary conditions imposed at $x=\pm L$ which satisfy \cite{Siegert}: $\left(\partial_x\Psi_n(\pm L)\mp ik_n\Psi_n(\pm L) \right)=0$.
Here we shall use atomic units for which $\hbar=m=1$. We choose to normalize the Siegert states such that: $2ik_n\int_{-L}^{L}\Psi_n^2(x)dx-\left[\Psi_n^2(L)+\Psi^2_n(-L)\right]=1$. For all but piece-wise potentials the solutions must converge with respect to $L$. One can categorize the Siegert solutions according to where they fall in the complex k-plane: Bound(anti-bound) solutions on the positive(negative) imaginary axis and resonance(anti-resonance) solutions on the fourth(third) quadrant. Using the relation $E_n=k_n^2/2$, the solutions' energy can be found.

Using the Siegert states solutions, the transmission amplitude for a one dimensional potential can be written as \cite{tolstikhinPRA1}: $t=2ke^{-2ikL}\sum_{n=1}^{2N} \frac{\Psi_n(L)\Psi_n(-L)}{k_n-k}$,
where the sum runs over all poles, i.e., bound, anti-bound, virtual, and resonance poles, and $k=\sqrt{2E}$. The sum above can be somewhat simplified by transforming it into a product. Such a product formula was first given by Tolstikhin \textit{et al.} in \cite{tolstikhinPRA1} for symmetric potentials and by Ostrovsky and Elander in \cite{ElanderPRA} for asymmetric potentials. For symmetric potentials, the product formula reads: $t=\frac{1}{2}e^{-2ikL}\left[\prod_{n=1}^{2N_+}\frac{k+k_n^+}{k-k_n^+}-\prod_{n=1}^{2N_-}\frac{k+k_n^-}{k-k_n^-}\right]
$,
where $N_+(N_-)$ correspond to symmetric(anti-symmetric) Siegert states. Since the poles $k_n$ are either purely imaginary or come in pairs $k_n$ and $-k_n^*$, the above product terms can each be written as a pure phase, the well known Blaschke factor \cite{nussenzveig-book}. Considering first the pairs of resonance and virtual states one can reformulate the product of each pair as: $\frac{k+k_n^\pm}{k-k_n^\pm}\frac{k-(k_n^\pm)^*}{k+(k_n^\pm)^*}=e^{2i\delta_n^\pm}$,
where after some trigonometric manipulations, one finds that $\delta_n^\pm=\arctan\left(\frac{k}{\textrm{Re}[k_n^\pm]}\frac{ (\Gamma_n^\pm/2)}{|E_n^\pm|-E}\right)$ with the common definition of the resonance energy: $E_{n}=\frac{k_n^2}{2}=\varepsilon_n-i\frac{\Gamma_n}{2}$. For the bound and anti-bound poles each term in the products read: $(k+k_n^\pm)/(k-k_n^\pm)=e^{2i\delta_n^\pm}$,
where for these poles $\delta_n^\pm=\arctan\left(\textrm{Im}[k_n]/k\right)$. Defining $\Delta_\pm=\sum_{n=1}^{N_\pm}\delta_n^\pm$ where the sum runs over all bound poles, anti-bound poles, and resonance-virtual pole pairs, the transmission amplitude can be brought to the following form: $t=ie^{-2ikL}e^{i(\Delta_++\Delta_-)}\sin(\Delta_+-\Delta_-)$,
and the transmission probability reads, $T=\sin^2(\Delta_+-\Delta_-)$.

\begin{figure}[htbp]
\includegraphics[width=\columnwidth,angle=0,scale=1,draft=false,keepaspectratio=true]{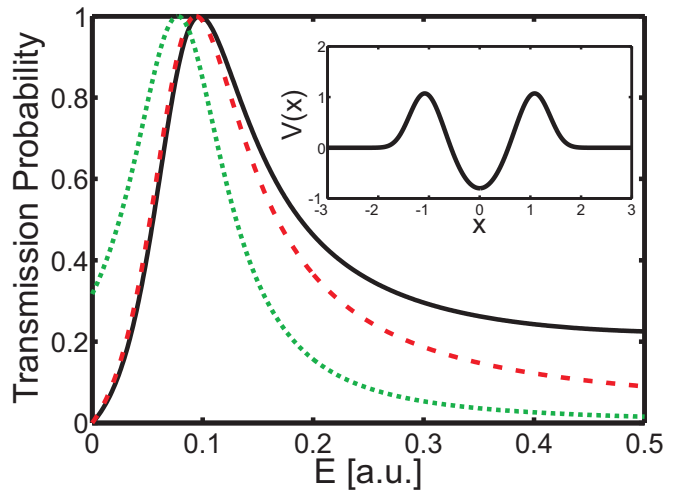}
\caption{(color online) The numerically exact transmission probability (solid(black) line) through a double barrier potential, see the text, using the following parameters: $\beta=5/2$, $\gamma=0.8$, and $\alpha=0.5$. We show only the first resonance peak. On top of the exact resonance peak we draw the BW resonance profile (dotted(green) line) (see Eq. \ref{eq:Breit-Wigner}) and the novel resonance profile given here (dashed(red)line) (see Eq. \ref{eq:one-res-trans}.)}
\label{Fig:ISR}
\end{figure}

Using the above  formula for the transmission probability, one can examine the contribution of a single resonance pole to the cross section. Without loss of generality, let us assume that the corresponding Siegert eigenstate is even. Then neglecting the contribution of all but a single resonance-virtual pole pair, the transmission probability reads,
\beq
\label{eq:one-res-trans}
T\approx\frac{(\frac{k}{k_r})^2(\Gamma/2)^2}{(E-|E_{res}|)^2+(\frac{k}{k_r})^2(\Gamma/2)^2}
\eeq
where  the resonance pole is located at $k_{res}=k_r-ik_i$, $E_{res}=\frac{k_{res}^2}{2}=\varepsilon-\frac{i}{2}\Gamma$, $k=\sqrt{2E}$, and we have used the trigonometric identity: $\sin(\arctan(x))=\frac{x}{\sqrt{1+x^2}}$. On first sight, Eq. \ref{eq:one-res-trans}, is very reminiscent of the BW resonance profile, see Eq. \ref{eq:Breit-Wigner}. There are, however, several crucial differences. First and foremost, the resonance peak has a maximum at the absolute value of the complex resonance pole energy. We have already demonstrated above, as shown in Fig. \ref{FIG::sqwell}, that this allows for the connection of each transmission unity with a single resonance pole even in the case of wide overlapping resonances. The resonance profile in Eq. \ref{eq:one-res-trans} also differs from the BW profile in that it is not a Lorentzian and is asymmetric with respect to its maximum, rising sharply to the left of the maximum and falling more slowly to the right of the maximum. Even though extension of the BW profile into an asymmetric one are known \cite{newton}, the current profile formula characterizes the asymmetry with only the resonance complex energy, i.e., with the same information needed to construct the BW profile, and \textit{does not require} the calculation of additional parameters.

Wide overlapping resonances are not the only case where the local description of resonance is abandoned for a non-local one. For resonances near the threshold, a similar belief that one cannot correlate between a resonance peak and a single pole exists. However, the resonance profile developed here shows that even very close to the threshold the resonance peak is due to a single resonance pole, and one can reconstruct the resonance peak in the cross section using only the resonance pole energy. Fig. \ref{Fig:ISR}, depicts the transmission probability near an isolated resonance in a double barrier potential of the form: $
V(x)=\left(\beta x^2-\gamma\right)e^{-\alpha x^4}$, see the inset. Clearly, the profile in Eq. \ref{eq:one-res-trans} reproduces the peak in the transmission probability very well using the same information used in the BW resonance profile which as seen in the figure fails to reproduce the resonance peak quantitatively. Thus one does not need to resort to non-local theories to describe resonance peaks near the threshold even when, as is in the case portrayed, the imaginary part and real part of the resonance energy are comparable in magnitude.

\begin{figure}[htbp]
\includegraphics[width=\columnwidth,angle=0,scale=1,draft=false,keepaspectratio=true]{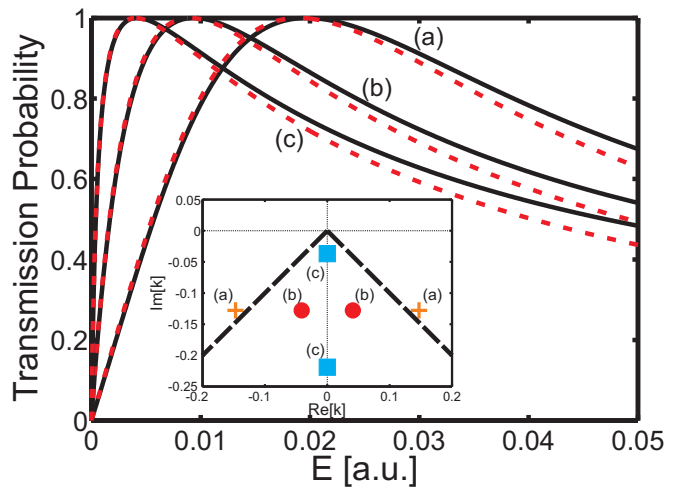}
\caption{(color online) The first resonance peak in the transmission probability for the double barrier potential, see the text, for different values of $\gamma$: $0.875$(a), $0.885$(b), and $0.89$(c). The numerically exact transmission (solid(black) line) is compared for each value of $\gamma$ with the novel resonance profiles presented here in Eqs. \ref{eq:one-res-trans} and \ref{eq:2AB-trans} (dashed(red) line). The inset portrays the poles of the scattering matrix closest to the imaginary k axis for each of the cases (a), (b), and (c). For convenience we also mark the $45$ degree bisector indicating in the third and fourth quadrants. Poles in the fourth(third) quadrant to the left(right) of the bisector have a negative position.}
\label{Fig:restrj}
\end{figure}

We turn now to discuss yet another important case where the local approximation has be erroneously discarded. In his seminal paper Nussenzveig \cite{nussenzveig} studied the effect of varying the potential on the poles of the scattering matrix. One of the most studied aspects has been the behavior of the poles as the potential is varied to support an additional bound state. It is well established that for a single s-wave like channel, looking in the k-plane, the resonance poles located in the fourth quadrant and the virtual state poles located in the third quadrant move toward the negative imaginary k-axis as the potential well is, for example, deepened. Eventually, two poles approaching from either side of the imaginary axis coalesce, thereafter, forming two anti-bound state poles which lie on the negative imaginary axis. Following the creation of two anti-bound poles, one of these poles moves up toward the origin and eventually leads to the formation of a new bound state, while the other moves down on the imaginary axis.

The connection between the motion of the resonance pole described above and the corresponding motion of the resonance peak in the cross section remains somewhat elusive even today. The main problem that arises in the analysis is that at some point, while approaching the negative imaginary axis, the resonance pole crosses the 45 degree bisector of the fourth quadrant of the k-plane making the resonance position negative. The resonance peak in the cross section, however, is still observable and is clearly situated above the threshold. Nussenzveig deemed these resonance poles with negative position as the only "meaningless" \cite{nussenzveig} poles. As we shall show they are just as meaningful as any other pole of the scattering matrix.

Consider the effect of deepening the well between the two barriers, see the potential depicted in Fig. \ref{Fig:ISR}, on the transmission spectrum. At some critical well depth, a new bound state appears. We wish to study the behavior of the resonance peak in the transmission spectrum as the potential depth is increased towards this critical value. Previous studies noted that as the well deepens the resonance peak approaches the origin. Figure \ref{Fig:restrj} portrays the first resonance peak in the transmission spectra for three different depths' of the potential well. As can be seen in the figure, the resonance peak moves toward the origin as we approach the critical value for which a new bound state is formed. Aside from the obvious shift in position, the different resonance peaks look almost the same. This is rather surprising since as we shall explain in the following, the potential responsible for the left most peak (marked (c)) doesn't support a resonance pole in the vicinity of the position of the peak.

Let us examine first the two peaks in figure \ref{Fig:restrj} marked (a) and (b). The potential responsible for peak (a) supports a resonance pole above the threshold similar to that depicted in figure \ref{Fig:ISR}. As can be seen the resonance profile given in Eq. \ref{eq:one-res-trans} (dashed(red) line) reproduces the numerically exact resonance peak in the spectrum also depicted (solid(black) line). For the peak marked (b), however, the potential is such that the resonance pole moved passed the 45-degree bisector of the fourth quadrant of the k-plane, see inset in figure \ref{Fig:restrj}. Therefore, the position of the resonances pole is now negative. As the figure shows, the profile in Eq. \ref{eq:one-res-trans} can be used very successfully to reproduce the resonance peak using only the resonance pole with the negative position. This puts the resonance poles with negative positions on the same footing as other resonance poles of the scattering matrix. The resonance pole's trajectory in the complex k-plane brings about yet another difficulty. As the pole approaches the negative imaginary axis, the resonance width defined as $\Gamma=2k_ik_r$ goes to zero since $k_r\rightarrow 0$. One might therefore conclude that such narrow resonances will be very hard to observe. Such a misconception stems from the expectation of a BW profile whose full width half maximum (FWHM) is $\Gamma$. The profile presented in Eq. \ref{eq:one-res-trans}, however, is asymmetric and has a FWHM of $\Gamma\sqrt{1+2\left(\frac{k_i}{k_r}\right)^2}$. For over the threshold, narrow resonances this is approximately $\Gamma$. However, for negative position resonances close to the negative imaginary axis one can no longer neglect the change in the FWHM. For the peak marked (b) in figure \ref{Fig:restrj}, the resonance pole width is of the order of $10^{-3}$ whereas the peak appearing in the figure has a FWHM of the order of $10^{-2}$. We reemphasize that the resonance peak, position and FWHM, are characterized using only the resonance pole complex energy, i.e., the same information used in the BW profile.

Further increase in the depth of the well, past the branch point where the two poles coalesced, produces two anti-bound states. Before proceeding, we point out that even though after the coalescence of the resonance and virtual poles, i.e., the resonance pole no longer exists, the peak is still apparent in the transmission spectrum. As figure \ref{Fig:restrj} clearly shows, the transmission spectrum seems somewhat oblivious to the \textit{identity crisis} the poles are experiencing. The change from resonance and virtual poles to two anti-bound states left no impression on the resulting spectrum. Since the colliding resonance and virtual states had the same spatial symmetry, the resulting two anti-bound states also share the same symmetry. We can therefore examine the transmission spectrum due to two anti-bound states of the same symmetry. Neglecting all contribution but those of the two anti-bound (AB) poles located at $k_1^{AB}=-ik_1^+$ and $k_2^{AB}=-ik_2^+$, the transmission probability reads
\begin{eqnarray}
\label{eq:2AB-trans}
T\approx\frac{E\left(k_1^++k_2^+\right)^2}{2\left(E-\frac{k_1^+k_2^+}{2}\right)^2+E\left(k_1^++k_2^+\right)^2}.
\end{eqnarray}
The remarkable agreement between the above  formula and the numerically exact transmission can be witnessed in Fig. \ref{Fig:restrj}.

In summary, we have demonstrated that a local theory of resonances can still be used even in cases previously thought impossible. Our analysis allows one to unequivocally connect the resonance peak in the transmission to specific poles of the scattering matrix throughout various changes of the scattering potential. Even for wide overlapping resonances one can connect each transmission unity with a single resonance pole.

We thank Dr. Ido Gilary and Prof. L. S. Cederbaum for helpful discussions. S.K. acknowledges the support of the Israel Ministry of Science, Culture and Sports. NM acknowledges the financial support of the ISF, grant number $96/07$.

\end{document}